\documentclass[letter]{aa}
\usepackage{graphicx}
\usepackage[varg]{txfonts}
\usepackage{units}
\usepackage{mathtools}
\usepackage{color}
\usepackage{natbib}
\usepackage{afterpage}

\newcommand{\beq}{\begin{equation}}
\newcommand{\eeq}{\end{equation}}

\begin{document}

\title{Anisotropic hydrodynamic turbulence in accretion disks}

\author{Moritz~H.~R.~Stoll\inst{1} \and
Wilhelm Kley\inst{1} \and
Giovanni Picogna\inst{1,2}}

\institute{
Institut f\"{u}r Astronomie und Astrophysik, Universit\"{a}t T\"{u}bingen,
Auf der Morgenstelle 10, D-72076 T\"{u}bingen, Germany\\
\email{\{moritz.stoll@, wilhelm.kley@\}uni-tuebingen.de}\\
\and
Universit\"{a}ts-Sternwarte, Ludwig-Maximilians-Universit\"{a}t München,
Scheinerstr. 1, D-81679 M\"{u}nchen, Germany\\
\email{picogna@usm.lmu.de}\\
}

\date{}

\abstract{Recently, the vertical shear instability (VSI) has become an attractive purely hydrodynamic 
candidate for the anomalous angular momentum transport required for weakly ionized accretion disks.
In direct three-dimensional numerical simulations of VSI turbulence in disks, 
a meridional circulation pattern was observed that is opposite to the usual viscous flow behavior.
Here, we investigate whether this feature can possibly be explained by an anisotropy of the VSI turbulence.
Using three-dimensional hydrodynamical simulations, we calculate the turbulent Reynolds stresses relevant for angular momentum
transport for a representative section of a disk.
 We find that the vertical stress is significantly stronger than the radial stress.
Using our results in viscous disk simulations with different viscosity coefficients for the radial and
vertical direction, we find good agreement with the VSI turbulence for the stresses and meridional flow; this 
provides additional evidence for the anisotropy.
The results are important with respect to the transport of small embedded particles in disks.}


\maketitle

\section{Introduction}
The exact origin of the driving mechanism of accretion disks is still not fully understood.
To accrete matter onto the central object, the matter needs to lose its angular momentum, and because
molecular viscosity is by many orders of magnitudes too small to facilitate the required
angular transport, it has been suggested that disks are driven by turbulence.
The discovery of a linear magneto-rotational instability (MRI) for rotating flows with a
negative angular velocity gradient has led to the suggestion that accretion disks are driven by 
magnetohydrodynamical (MHD) turbulence \citep{1998RvMP...70....1B}. 
However, the MRI only works efficiently for well-ionized media,
for example, in disks around compact objects, but for lower ionization levels the non-ideal MHD effects
become stronger and the operability of the MRI questionable. In particular for the cool and low-ionized 
regions, so-called {\it \textup{dead zones}} with very low turbulent activity have been predicted \citep{1996ApJ...457..355G}.

Hence, other alternatives are sought for. In the past years the purely hydrodynamical vertical shear instability (VSI)
has attracted some attention because the only requirement is a vertical shear in the angular velocity profile,
which is in fact a natural consequence of a radial temperature gradient in the disk, for example, induced by irradiation from the central object.
Through numerical simulations and linear analysis it has been shown that the
VSI operates efficiently for vertically isothermal disks \citep{Nelson2013} as well as for 
fully radiative disks that include stellar irradiation \citep{Stoll2014A&A...572A..77S}.
This makes the VSI a promising candidate to bring at least some life back into the dead zones.
For many astrophysical applications it is useful to parameterize the turbulence and describe the angular momentum
transport by an effective viscous prescription, for example, the well-known ansatz by \citet{1973A&A....24..337S}
where the uncertainties of the turbulence are summarized in one constant parameter $\alpha$ \citep{1999ApJ...521..650B}.
Such an approach using one parameter is applicable for an isotropic turbulence, and useful when the interest is in the overall radial
evolution of the disk. To analyze the internal flow field of the disk, which is important for the motion of
embedded small particles, it is necessary to take possible non-isotropy effects into account.

Here, we demonstrate 
that the flow reversal found in recent VSI-turbulent disks \citep{2016A&A...594A..57S} can in fact be traced back to
the intrinsic anisotropy of the VSI-turbulence. 
Using multidimensional hydrodynamical simulations, we calculate the effective radial and vertical transport coefficients and use them in viscous disk simulations.
 
\section{Model setup}\label{sec:setup}
We use the \textsc{pluto} code \citep{Mignone2007} for our simulations, where we model a section of a locally isothermal accretion disk.
For the direct turbulent simulations we use a full 3D setup and spherical coordinates ($R$,$\theta$,$\phi$), 
while for comparison laminar viscous simulations ($\alpha$-disk), we use a 2D axisymmetric setup and cylindrical coordinates ($r, z$).
The model parameters are given in Table~\ref{Tab:paras}.
\begin{table}[t]
    \caption{Model parameter: Domain size and grid resolution}
    \label{Tab:paras}
    \centering
    \begin{tabular}{l c c}
        \hline\hline
        Parameter & $\alpha$-model & VSI\\
        \hline
        Radial range [$5.2\ \mbox{au}$] & $0.4$ - $2.5$ & $0.4$ - $2.5$ \\
        Vertical range [H] & $\pm 4$ & $ \pm 5$\\
        Phi range [rad] & 2D & $0$ - $2\ \pi$ \\
        Radial grid size & $600$  & $600$\\
        Theta grid size & $121$  & $128$\\
        Phi grid size & 2D  & $1024$ \\
        \hline\hline
    \end{tabular}
\end{table}
Even though some of the simulations are performed in spherical coordinates, the results are analyzed in cylindrical coordinates.
The initial disk is axisymmetric and extends from $0.4\ r_\mathrm{0}$ to $2.5\ r_\mathrm{0}$, where $r_0 = \unit[5.2]{au}$.
The initial density profile is given by vertical hydrostatic equilibrium
\begin{equation}
    \label{eq:surfprof}
    \rho(r,z)=\rho_0\, \left(\frac{r}{r_0}\right)^{p}\, \exp{\left[\frac{GM}{c_\mathrm{s}^2}\left(\frac{1}{R}-\frac{1}{r}\right)\right]}\,,
\end{equation}
where $\rho_0$ is the gas mid-plane density at $r = r_0$, and $p=-1.5$.
In our locally isothermal approximation the temperature of the disk is a function of the cylindrical radius only,
\begin{equation}
    \label{eq:tempprof}
    T(r) = T_0\, \left(\frac{r}{r_0}\right)^{q} \,,
\end{equation}
where we choose $q=-1,$ which causes the disk aspect ratio to
be constant and $T_0$ such that $H/r =h= 0.05$.
The pressure is given by $P = c_s^2 \rho$, where $c_s$ is the isothermal sound speed with
$c_s \propto r^{-1/2}$.
The gas moves initially with the angular velocity given by the Keplerian value, corrected by the pressure support \citep{Nelson2013}
\begin{equation}
  \Omega(r,z) = \Omega_K \left[ (p +q)\left( \frac{H}{r} \right)^2 + (1+q) - \frac{q r}{\sqrt{r^2 + z^2}} \right]^{\frac{1}{2}} \,,
  \label{eq:omega}
\end{equation}
with $\Omega^2_K(r) = G M_*/r^3$, and the meridional flow ($u_r, u_z$) is set to zero.
At the inner and outer boundary we use reflecting boundaries. To increase numerical stability for the
 turbulent runs, we damp the variables $\rho, u_r, u_z$
 to the initial values on a timescale of half a local orbit following the recipe by \citet{DeVal-Borro2006}, 
 with a damping applied at the inner boundary from \unit[0.4-0.5]{$r_0$} and at the outer boundary from \unit[2.3-2.5]{$r_0$}.
 At the vertical boundaries we use reflective boundaries when the flow is directed inward and zero gradient otherwise.
 The VSI model is inviscid, and for the $\alpha$-disk model the kinematic viscosity is given by $\nu = 2/3 \alpha c_s H$,
 with a constant $\alpha = 5 \cdot 10^{-4}$, which matches the outcome of VSI model.

\section{Disk structure}
We compare the structure of the VSI unstable disk to viscous
disks described by an $\alpha$-parameter, and study the main differences. The VSI disk is evolved to a quasi-equilibrium state before the analysis is performed.
We focus in particular on the flow field and stress tensor, and refer to \citet{Nelson2013} 
and \citet{Stoll2014A&A...572A..77S,2016A&A...594A..57S} for an analysis of the turbulent flow structure.

\subsection{Radial velocity}
\begin{figure}[tb]
    \centering
    \includegraphics{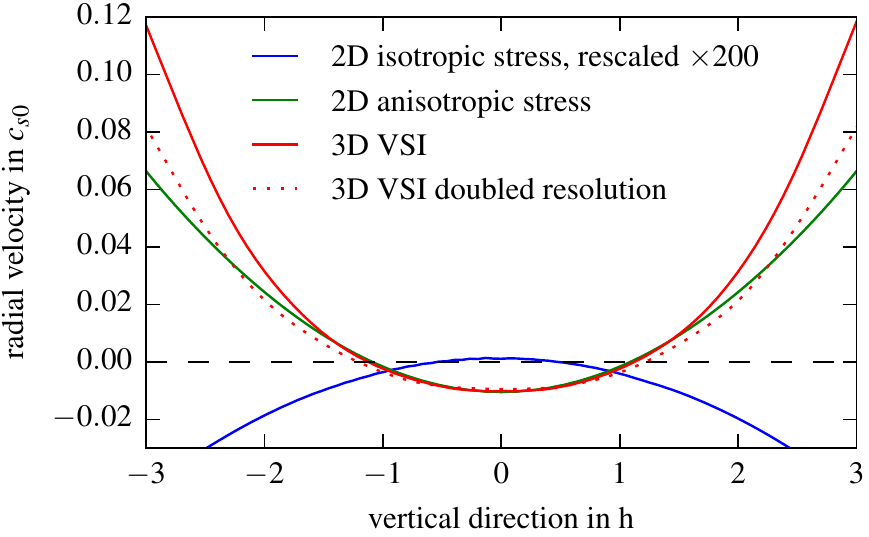}
    \caption{Radial velocity averaged over 50 orbits. 
 We compare the disk with alpha-viscosity ($\alpha=5 \cdot 10^{-4}$, blue curve) to 
 a disk with active VSI (red curve), a disk with active VSI and doubled resolution (red dotted curve) and a viscous disk with anisotropic stress similar to the VSI disk (green curve, details see Sect.~\ref{sec:visc}).  
 For the turbulent disk the velocity has been azimuthally averaged. 
 The profile shown is in units of sound speed $c_{s0}$ and at $r=r_0$. The viscous case has been rescaled to better visualize the difference.}
    \label{fig:radialVelocity}
\end{figure}

While the density distributions of the viscous and turbulent disks in equilibrium are very similar due to the necessary pressure equilibrium,
there is an important difference in the meridional flow within the disk in particular for the radial velocity. 
We show the azimuthally averaged radial velocity $u_r$ at $r=r_0$ as a function of the vertical distance for the turbulent and the standard viscous disk in Fig.~\ref{fig:radialVelocity}. For the turbulent disk we averaged the velocity in time over 50 orbits and in space around $r_0$ in the region $(0.8 - 1.25) r_0$. For better visibility we rescaled the viscous case by a factor of $200$. Obviously, the two radial velocity profiles have an opposite behavior. The standard
viscous disk using a single (isotropic) value of $\alpha$ shows the typical outflow in the midplane that has been predicted analytically by \citet{1984SvA....28...50U} and was later shown in fully time-dependent numerical simulations \citep{1992ApJ...397..600K}. This behavior of $u_r(z)$ can be derived from the equilibrium angular momentum equation that contains the vertical disk structure \citep{1984SvA....28...50U}. In spite of the outwardly directed flow in the midplane, the total vertically integrated mass flow is nevertheless directed inward in case of accreting disks \citep{1992ApJ...397..600K}. 

In contrast, the mean flow field for the turbulent flow (labeled 3D VSI and shown also with a dotted line for a model
with double resolution in all directions) is fully reversed; it is not only negative in the midplane and positive in the corona, but also much stronger, as indicated by the different scaling of the curves in Fig.~\ref{fig:radialVelocity}.
This special feature of the meridional flow field in the VSI case has been found for isothermal as well as radiative disks in \citet{2016A&A...594A..57S},
but was not analyzed with respect to anisotropic turbulence.
From the direct comparison to the viscous case, it is clear that with a standard shear viscosity prescription using a constant $\alpha$-value or a constant kinematic viscosity, no agreement can be obtained because this will always lead to an inverted parabolic type of profile \citep{2013A&A...551A..75J}. This raises the general question whether the VSI turbulence in disks can be described by a standard Navier-Stokes approach to model the angular momentum diffusion. In Sect.~\ref{sec:visc} we show that we obtain a good match of a fully turbulent and viscous flow for a non-anisotropic turbulent viscosity where the radial and vertical parts enter with a different strength (see curve 2D anisotropic stress),
which might be expected for the clearly non-isotropic character of the flow structure in VSI turbulence \citep{Stoll2014A&A...572A..77S}.

\subsection{Turbulent stresses}
To analyze the effect of the turbulence with respect to angular momentum transport in the disk, it is necessary to calculate the turbulent stresses of the VSI disk. For the overall mass flow in accretion disks, which is driven by outward angular momentum transport, the $r\phi$-component of the Reynolds stress tensor, $R$, is the most important component because it is generated by the strong shear in the azimuthal velocity. In the case of VSI turbulent disks, it is clear that the vertical dependence of the turbulent stresses ($z\phi$-component) may be of importance as well.  Hence, from our simulations we calculate the following turbulent Reynolds stresses
\begin{equation}
    R_{i,\phi} (z) \, = \, <\rho u_i \delta u_{\phi}>_{t, \phi, r} \,,
    \label{eq:stresses}
\end{equation}
where $u_i$ with $i$ in $(r,z)$ denote the radial and vertical flow velocity ($u_r, u_z$) and $\delta u_{\phi}$ is the deviation of the azimuthal velocity, $u_\phi = r \Omega$, from unperturbed equilibrium, as given by Eq.~(\ref{eq:omega}).  For the later analysis it is beneficial that the stresses are calculated in cylindrical coordinates ($r,z$), see Sect. \ref{sec:visc}.  Since we are interested in particular in the vertical dependence of the stresses, $R_{i,\phi}$ is calculated by averaging over the full azimuth ($2 \pi$), over a small radial range around a reference radius, and over time.  Here, we time-average from orbit 80 to orbit 200 using a series of over 60 snap shots and space-average in radius from \unit[0.75 - 1.35]{$r_0$}. 

\begin{figure}[tb]
    \centering
    \includegraphics{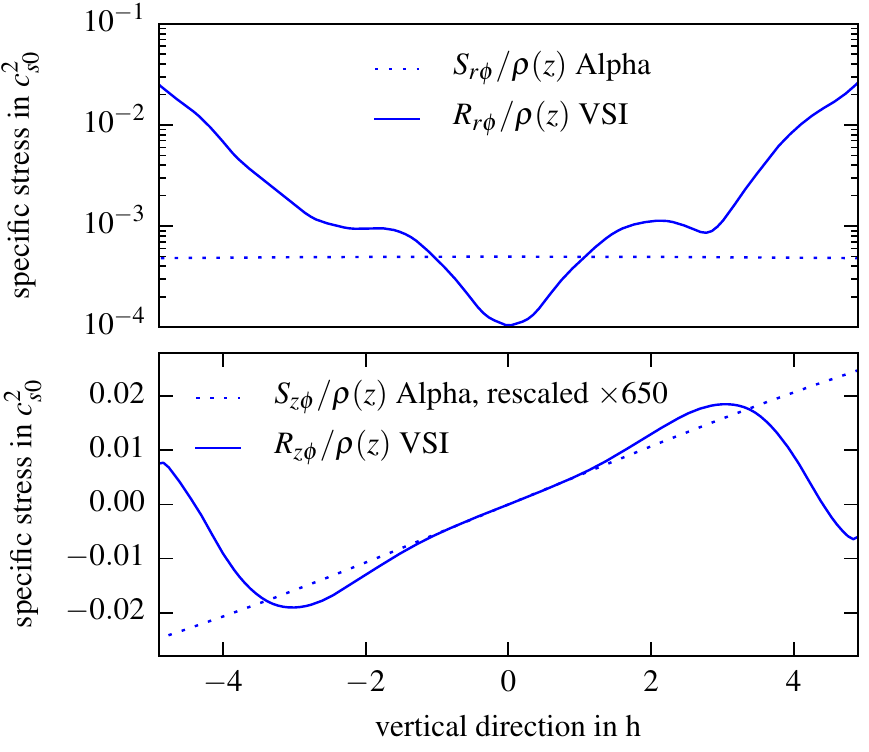}
    \caption{Specific stress tensor (stress per density) in units of $c_{s0}^2$. The solid lines represent the Reynolds stress ($R$) of the VSI simulation 
    and the dotted lines refer to the viscous stress tensor ($S$) using $\alpha = 5 \cdot 10^{-4}$. 
 The $S_{z\phi}$ line was rescaled by a factor of 650 to match to the VSI model. 
  The dotted line in the upper panel has a constant value of $5\cdot 10^{-4}$.}
    \label{fig:stresses}
\end{figure}

The results of this averaging procedure are shown in Fig.~\ref{fig:stresses}. The solid curves refer to the specific stresses $ R_{i,\phi}(z) / \rho(r_0,z)$ ($i=r,z$) 
in units of $c_{s0}^2$, where $\rho(r_0,z)$ is the equilibrium density distribution, given by Eq.~(\ref{eq:surfprof}), 
and $c_{s0}$ is the sound speed, both at the reference radius $r_0$.

We compare it to the viscous shear stress prescription (see Eq.~\ref{eq:stresscyl} with $\nu = \nu_r = \nu_z$) using $\alpha = 5 \cdot 10^{-4}$ (dotted lines),
which is close to  the average of the $r \phi$-component of the specific VSI stress, $4 \cdot 10^{-4}$. 
We rescaled the $z \phi$-component of the viscous shear stress by a factor of 650 to match the value from the VSI stress. 
From this large rescaling factor it is immediately clear that $R_{z\phi}$ is far larger than expected from an isotropic viscous shear prescription.
We note that the deviation of the vertical $R_{r\phi}/\rho$-profile from the constant $S_{r\phi}/\rho$-profile is not important for our argument, which is why for simplicity we chose a constant $\alpha$.
For the angular momentum transport only the vertical average of this stress component plays a role.
In our case, the very large $z\phi$-component of the stress dominates the meridional flow and the influence of $R_{r\phi}$ is small.

\section{Anisotropic viscosity of the VSI turbulence}
\label{sec:visc}

From our numerical studies of the VSI turbulence, in particular the vertical component of the Reynolds stress tensor,
we can infer that the turbulence is non-isotropic.
Even though we have integrated the hydrodynamical equations using spherical coordinates,
we calculated and display $R_{i\phi}$ in cylindrical coordinates because this is simpler to analyze, as we explain in
the following. In our approach we investigate whether the turbulent Reynolds stresses, $R$, can be modeled by a viscous ansatz 
where $R$ is replaced by the standard viscous stress tensor, $S$, with an effective turbulent eddy viscosity as introduced by
Boussinesq.
In cylindrical coordinates the change in angular momentum is given by the following
evolution equation:
\beq
 \label{eq:angmom}
   \frac{\partial \rho r^2 \Omega}{\partial t}   +
   \nabla \cdot ( \vec{u} \rho r^2 \Omega) =
  - \frac{ \partial P}{\partial \phi}
  +  \nabla \cdot ( r \vec{S}_\phi ) \,.
\eeq
For an axisymmetric flow, that is, when $\partial / \partial \phi$ vanishes, 
the vector $\vec{S}_\phi$ (the $\phi$-row of the viscous stress tensor) is given by
\citep{1978trs..book.....T}
\beq
\label{eq:stresscyl}
  \vec{S}_\phi = ( S_{r\phi}, S_{z\phi})
  =  \left( \, \rho \nu_r r \frac{\partial \Omega}{\partial r}, \,
       \rho \nu_z r \frac{\partial \Omega}{\partial z} \, \right) \,.
\eeq
Axisymmetry is expected for accretion disk flows, and we found this in our simulations.
In Eq.(\ref{eq:stresscyl}) we allowed for the option of an anisotropic viscosity by splitting the
kinematic viscosity into two components $\nu_r$ and $\nu_z,$ where $\nu_r$ refers to the radial part,
which typically is the main contribution in accretion disks in driving the angular momentum transport and
mass accretion. The $\nu_z$ part connects to the vertical variation of the angular velocity $\Omega$ and
has commonly been assumed to be on the same order as $\nu_r$. Indeed, using $\nu_z = \nu_r$ and performing viscous
axisymmetric, two-dimensional simulations, we find the typical meridional flow field in disks with outflow in the midplane
and inflow in upper layers of the disk, as shown in Fig.~\ref{fig:radialVelocity}
by the curve labeled 'isotropic stress', as
found in classic studies \citep{1984SvA....28...50U,1992ApJ...397..600K}.
The outflow in the midplane can easily be derived by an analysis of the radial flow
that can be obtained from the angular momentum equation \citep[see also][]{2011A&A...534A.107F,2013A&A...551A..75J}.
From Eq.~(\ref{eq:angmom}) we find for the equilibrium state that
\beq
\label{eq:ur-eq}
  u_r  \propto  \frac{1} {\rho r^2 \Omega} \,
   \left(  \frac{1}{r} \frac{\partial (r^2 S_{r \phi})}{\partial r}
         +  \frac{\partial r S_{z \phi}}{\partial z} \right) \,.
\eeq
Expanding Eq.~(\ref{eq:omega}) around the midplane (small $z$), we find
\beq
\Omega(r,z) = \Omega_K(r) \, \left( 1 + \frac{q}{4} \frac{z^2}{r^2} + \frac{q + p}{2} h^2 \right) \,,
\eeq
where $q$ and $p$ are the exponents in the radial power laws of the density and temperature, respectively,
and $h = H/r$ denotes the relative scale height of the disk.
Using this relation for $\Omega(r,z),$ we find
\beq
       \frac{\partial \Omega}{\partial z} = \Omega_K \, \frac{q}{2} \, \frac{z}{r^2}
\eeq
and then, neglecting terms of order $z^2/r^2$
\beq
S_{z\phi} =  q \, \rho \nu_z \, \frac{\Omega_K}{2} \frac{z}{r}
    \quad  \quad \mbox{and}  \quad  \quad 
S_{r\phi} = - \rho \nu_r \, \frac{3}{2} \Omega_K \,.\eeq
Combining this with the $S_{z\phi}$ relation, we find
\beq
\label{eq:szp}
S_{z\phi} =  - \frac{q}{3} S_{r \phi} \, \frac{\nu_z}{\nu_r} \, \frac{z}{r} \,.
\eeq
This is plotted in the lower panel in Fig.~\ref{fig:stresses} using $\nu_z = 650 \nu_r$,
which demonstrates that the linear dependence of the specific $S_{z\phi}$-stress is a direct consequence
of Eq.~(\ref{eq:szp}).
From Eq.~(\ref{eq:ur-eq}) we find for the relation determining the sign of $u_r$ in the midplane the
following relation:
\begin{eqnarray}
u_r (z=0)  & \propto &   \left(  2 S_{r\phi}  + r \frac{\partial S_{r\phi}}{\partial r}  +  r \frac{\partial S_{z\phi}}{\partial z} \right)
\nonumber \\
 & \propto & \rho \, \left[ \nu_r \left( -  3  - \frac{3}{2} ( q + p + \frac{3}{2} ) + \frac{9}{4} \right)  + \frac{q}{2} \nu_z \right]
,\end{eqnarray}
where for the last step we assumed that the viscosity has an $\alpha$-type behavior with $\nu \sim \alpha c_s^2 / \Omega_K$
and $\rho \propto r^p$ and $c_s^2 \propto r^q$. We note that the direction of flow in the disk midplane
can be influenced by the slopes ($p,q$) in the disk stratification \citep{2011A&A...534A.107F,2017arXiv170101912P}.
From the above relation we find directly that for our disk with slopes $q=-1$ and $p=-3/2$ that 
\beq
u_r (z=0) \propto  \left[  \frac{3}{2}  \alpha_r   -  \alpha_z \right]  \,.
\eeq
For isotropic turbulence with $\alpha_r = \alpha_z$
we therefore have outflow in the disk midplane. For the turning point we find that $\alpha_z$ must
be larger than $1.5 \alpha_r$.
Upon increasing $\alpha_z$ over $\alpha_r,$ the midplane radial velocity becomes more and more negative,
and the entire vertical flow profile eventually reverses.
In our numerical simulations of viscous disks using different values for $\nu_z$ we could indeed
find the observed flow reversal for moderate values of $\nu_z/\nu_r$.
By increasing $\nu_z$ further, we found that the VSI turbulence can be modeled by an anisotropic eddy viscosity with 
$\nu_z$ over two magnitudes larger than $\nu_r$ (650 for our chosen parameter, as shown in Fig.~\ref{fig:stresses}).

In performing the comparison simulations, we initially tried to use the same spherical coordinate system
that we used for the turbulent VSI simulations by just increasing $\nu_\theta$ over $\nu_R$
in the corresponding components of the stress tensor in spherical coordinates.
However, this did not lead to the expected results, in particular, the inversion of the parabola of $u_r(z)$.
The results displayed in Fig.~\ref{fig:radialVelocity} were therefore
obtained using a cylindrical coordinate system.
Because the usage of a spherical coordinate system is beneficial in disk simulations, the equations need
to be transformed from $r,z$ to $R, \theta$.

This means we need to transform the viscous part, $\nabla \cdot ( r \vec{S}_\phi$), to spherical coordinates.
The transformation equations are
\beq
r = R \sin \theta   \quad \quad   \text{and} \quad \quad  z = R \cos \theta
,\eeq
and the components of the stress tensor transform according to
\begin{eqnarray}
\label{eq:SRP}
S_{R\phi}  & = & S_{r\phi} \sin \theta  + S_{z\phi} \cos \theta  \,, \\
S_{\theta\phi} & = &  S_{r\phi}  \cos \theta  - S_{z\phi} \sin \theta \,.
\label{eq:STP}
\end{eqnarray}
The derivatives of $\Omega$ transform then as
\begin{eqnarray}
\label{eq:domr}
\frac{\partial \Omega}{\partial r} & = & \sin \theta \frac{\partial \Omega}{\partial R}
                                  \, + \, \frac{\cos \theta}{R} \frac{\partial \Omega}{\partial \theta} \\
\label{eq:domz}
\frac{\partial \Omega}{\partial z} & = & \cos \theta \frac{\partial \Omega}{\partial R}
                                  \, - \, \frac{\sin \theta}{R} \frac{\partial \Omega}{\partial \theta}
.\end{eqnarray}
Using Eq.~(\ref{eq:stresscyl}) in Eqs.~(\ref{eq:SRP}) and (\ref{eq:STP}) and substituting this into
Eqs.~(\ref{eq:domr}) and (\ref{eq:domz}), we obtain for the stress tensor components in spherical coordinates
\begin{eqnarray}
S_{R\phi}  & = &  \rho R \sin \theta \left[ \,  \left( \nu_r \sin^2 \theta
                   + \nu_z \cos^2 \theta \right) \,  \frac{\partial \Omega}{\partial R} \right.
   \label{eq:SRPfull} \nonumber  \\
         & + &  \left. (  \nu_r - \nu_z ) \, \frac{\cos \theta \sin \theta}{R} \, \frac{\partial \Omega}{\partial \theta} \, \right] \\
S_{\theta\phi}  & = &  \rho R \sin \theta \left[ \, \sin \theta \cos \theta \,
   \left(  \nu_r - \nu_z \right) \, \frac{\partial \Omega}{\partial R} \, \right.
   \label{eq:STPfull} \nonumber  \\
& + &  \left. \left( \nu_r \cos^2 \theta + \nu_z \sin^2 \theta \right) \, \frac{1}{R} \frac{\partial \Omega}{\partial \theta} \right]
.\end{eqnarray}
As can be inferred from these equations, the relatively simple anisotropic relation in cylindrical coordinates
leads  to complex coupled equations in spherical coordinates with cross terms in the derivatives
for $\Omega$.
For the isotropic case we can set $\nu_r = \nu_z = \nu$ and obtain the standard relations for the stress tensor components
in $R \theta$-coordinates \citep{1978trs..book.....T}.
Now it becomes clear why merely increasing the $\theta$ part of the viscosity in $R \theta$-coordinates did not yield the correct answer when compared to the VSI case.  Using the full viscosity terms as described in Eqs.~(\ref{eq:SRPfull}) and (\ref{eq:STPfull})
in the numerical simulations yields the correct behavior. However, this comes with a serious drawback because
in this case the numerical integration required much smaller time steps for numerical stability than the cylindrical simulation with the
same viscosity coefficients. 
Performing multidimensional simulations
of viscous disks mimicking the non-isotropic behavior of the VSI turbulence in $R \theta$-coordinates therefore
requires a considerable numerical effort. One solution to this problem may be the usage of an implicit solver
for the viscous terms.

\section{Conclusions}
From direct three-dimensional simulations of locally isothermal accretion disks we observed that the eddies introduced by the VSI generate stresses
that are strongly anisotropic. Specifically, we found that the vertical $z\phi$-component of the Reynolds stress is enhanced by a factor of 650 over 
the standard $r\phi$-part. By performing viscous disk simulations using a non-isotropic viscosity with $\alpha_z$ highly enlarged over $\alpha_r,$ we could obtain the same flow reversal as seen in the VSI disk, which verifies the non-isotropy of the viscosity. 
Hence, the reversal of the radial flow profile compared to the usual $\alpha$-model is a clear consequence of the anisotropy.
This will have an effect on the dust migration processes \citep{2016A&A...594A..57S} that need to be contrasted to the outward drift in the midplane
viscous models \citep{Takeuchi2002ApJ...581.1344T}. 
From this we can conclude that we need to be careful with turbulence models imposed on accretions disks when we adopt  viscous models to describe them.

The meridional circulation of MRI-turbulent disks has been analyzed by \citet{2011A&A...534A.107F}, who found a similar mean flow dynamics in the disk 
but did not attribute it to a non-isotropic turbulence (with large $S_{z\phi}$) but rather to a radial variation in the magnitude of the viscosity.
However, their turbulent $R_{z\phi}$ profile (in Fig. 5) indicates that it may be enhanced over the standard viscous value as well,
which could also be a reason for the flow reversal.

Even though the analysis is performed for simplicity for a locally isothermal disk, our results are quite general as
simulations for fully radiative disks 
show  the same behavior although they have different radial and vertical temperature profiles.
For this purpose, we reanalyzed our radiative simulations in \citet{2016A&A...594A..57S} and found a similar anisotropy.
Concerning numerical resolution, simulations with double resolution show the same results as shown in Fig.~\ref{fig:radialVelocity}.
Nevertheless, further exploration needs to be done in order to check how the anisotropy factor varies for different disk parameters and to explore the possibility
of anisotropic stresses in MRI models.

\begin{acknowledgements}
    Moritz Stoll received financial support from the Landesgraduiertenf\"orderung of the state of Baden-W\"urttemberg and through the Ger    man Research Foundation (DFG) grant KL 650/16.
    G. Picogna acknowledges the support through 
    DFG-grant KL 650/21 within the collaborative research program ``The first
    10 Million Years of the Solar System''. 
    Some simulations were performed on the bwGRiD cluster in T\"ubingen, funded
    by the state of Baden-W\"urttemberg and the DFG.
    We thank Roman Rafikov for providing us with a preprint and very useful discussions.
\end{acknowledgements}

\bibliographystyle{aa}
\bibliography{calibre,biblio,wk}{}

\begin{appendix}

\end{appendix}

\end{document}